\documentclass[prd,aps,showpacs,nofootinbib]{revtex4}
\usepackage{amssymb}
\usepackage{graphicx}
\usepackage{amsmath}

\newcommand{\be}{\begin{equation}}
\newcommand{\ee}{\end{equation}}
\newcommand{\bq}{\begin{eqnarray}}
\newcommand{\eq}{\end{eqnarray}}

\begin{document}
\title{Lorentz violation with an invariant minimum speed as foundation of the Gravitational Bose Einstein Condensate of a Dark Energy Star} 
\author{**Cl\'audio Nassif Cruz, *Rodrigo Francisco dos Santos and A. C. Amaro de Faria Jr.}
\affiliation{\small{**CPFT: Centro de Pesquisas em F\'isica Te\'orica, Rua Rio de Janeiro
1186/s.1304, Lourdes, 30.160-041, Belo Horizonte-MG, Brazil.\\
 *UFF: Universidade Federal Fluminense, Instituto de F\'isica, Av. Gal. Milton Tavares de Souza s/nº, Gragoat\'a, 24.210-346, Niter\'oi-RJ, Brazil.\\
 IEAv: Instituto de Estudos Avan\c{c}ados, Rodovia dos Tamoios Km 099, 12220-000, S\~ao Jos\'e dos Campos-SP, Brazil.\\
 **claudionassif@yahoo.com.br, *santosst1@gmail.com, antoniocarlos@ieav.cta.br}} 

\begin{abstract}
We aim to search for the connection between the spacetime with an invariant minimum speed so-called Symmetrical Special Relativity (SSR) with Lorentz violation and the Gravitational Bose Einstein Condensate (GBEC) as the central core of a star of gravitational vacuum (gravastar), where one normally introduces a cosmological constant for representing an anti-gravity. This usual model of gravastar with an equation of state (EOS) for vacuum energy inside the core will be generalized for many modes of vacuum (dark energy star) in order to circumvent the embarrassment generated by the horizon singularity as the final stage of a gravitational collapse. In the place of the problem of a singularity of an event horizon, we introduce a phase transition between gravity and anti-gravity before reaching the Schwarzschild (divergent) radius $R_S$ for a given coexistence radius $R_{coexistence}$ slightly larger than $R_S$ and slightly smaller than the core radius $R_{core}$ of GBEC, where the metric of the repulsive sector (core of GBEC) would diverge for $r=R_{core}$, so that for such a given radius of phase coexistence $R_S<R_{coexistence}<R_{core}$, both divergences at $R_S$ of Schwarzschild metric (a fine shell of baryonic matter involving the core of GBEC) and at $R_{core}$ of the repulsive core are eliminated, thus preventing the formation of the event horizon. So the causal structure of SSR helps us to elucidate such puzzle of singularity of event horizon by also providing a quantum interpretation for GBEC and thus by explaining the origin of a strong anisotropy due to the minimum speed (dark cone) that leads to the phase transition gravity/anti-gravity during the collapse of the star. Furthermore, due to the absence of an event horizon of black hole (BH) where any signal cannot propagate, the new collapsed structure presents a signal propagation in its region of coexistence of phases where the coexistence metric does not diverge, thus leading to emission of radiation.  
\end{abstract}
\pacs{11.30.Qc,98.80.Qc,04.20.Dw,04.20.Jb,04.70.Bw\\
 Keywords: quantum black hole, dS-space, dark energy star, vacuum energy, gravitational Bose-Einstein condensate, minimum speed, Planck length, phase transition, Lorentz violation.}
\maketitle

\section{Introduction}

The search for understanding the cosmological vacuum has been the issue of hard investigations\cite{1}. Several models have been suggested\cite{2} in order to postulate the existence of a quantum vacuum (e.g: fluid of Zeldovich), but without proposing a physical interpretation for the vacuum until the emergence of a new kind of Deformed Special Relativity (DSR) so-called Symmetrical Special Relativity (SSR)\cite{3}\cite{4}\cite{5}\cite{6}, which has brought a new interpretation for the quantum vacuum by means of the concept of an invariant minimum speed $V$ directly related to the miminum length (Planck length $L_P$), i.e., $V\propto L_P$\cite{3}, which changes the causal structure of spacetime and geometrizes the quantum phenomena\cite{5}\cite{uncertainty}. This minimum speed modifies the Minkowski metric of the spacetime by providing the SSR-metric\cite{3}, which is a kind of conformal metric as is the de-Sitter (dS)-metric\cite{Rodrigo}. Such 
SSR-metric is able to represent the metric of the Gravitational Bose Einstein Condensate (GBEC), which represents the core of a gravastar\cite{7}\cite{8}\cite{9}\cite{10}\cite{16}\cite{17}
\cite{18}\cite{19}\cite{20}. So we are able to map the SSR-metric into the GBEC-metric in such a way that we can associate the cosmological constant of GBEC with the minimum speed connected to the vacuum energy density\cite{3}. Thus by linking the star structure equations with the causal structure of spacetime of SSR, a phase 
transition appears in place of the event horizon postulated by Chapline and others\cite{11}\cite{12}\cite{14}\cite{15}. Such a phase transition can be related to this new causal structure of SSR, where SSR describes perfectly a fluid, which is similar to a relativistic superfluid of type of cosmological fluid, being the constituent of GBEC.

In the 2nd. section, we make a brief review of the concept of an invariant minimum speed in spacetime and its implications in the existence of the cosmological constant\cite{3}, basing on the Dirac's Large Number 
Hypothesis (LNH). Thus we present a conformal metric of spacetime due to the presence of the minimum speed and so we show the cosmological fluid (vacuum energy) or its equation of state (EOS) generated by Symmetrical Special Relativity (SSR).

In the 3rd. section, we introduce the GBEC metric of a gravastar. We make a study about the 
well-known concept of anisotropy inside the GBEC, which permits the emergence of the phase transition (gravity/anti-gravity), which appears in place of the event horizon during the gravitational collapse. Such phase 
transition is an implication of a more general equation of state (EOS) for a real structure of GBEC with many modes of vibrational states instead of a unique EOS of vacuum ($p=-\rho$ with $w=-1$), so that the anisotropy could be better understood for explaining the null pressure ($p=0$ with $w=0$) given in the region of phase coexistence (gravity/anti-gravity). 

In the section $4$, we map the well-known dS-metric that governs GBEC into the SSR-metric, by showing an interesting similarity between them, as they can be written in the same form with equivalent conformal 
factors, where there is a relationship between the GBEC (core) radius and the minimum speed for the core radius. 

Finally, in the section $5$, we investigate deeper what occurs in the region of phase transition, i.e., at the coexistence radius ($R_{coexistence}$) of the two phases. We aim to obtain the spacetime metric in such a region, where we verify that the metric does not diverge as occurs in the classical case for the Schwarzchild radius $R_S$ (no phase transition). We show that there is no divergence in such a region of phase transition due to the fact that the coexistence radius of phases ($R_{coexistence}$) is slightly larger than the Schwarzchild radius by preventing the singularity of event horizon, i.e., we must find $R_{coexistence}>R_S$.

\section{A brief review of Symmetrical Special Relativity (SSR): spacetime transformations 
with a universal minimum speed emerging from Dirac's Large Number Hypothesis}

Let us first show the need of emergence of a universal minimum speed as a new fundamental
constant of nature, according to a careful analysis of Dirac's Large Number Hypothesis (LNH). Such a universal minimum speed $V$ has the same status of the invariance of the speed of light 
($c$), however $V$ is given for lower energies related to gravity, which is the weakest 
interaction, whereas $c$ is well-known as being associated with the electromagnetic fields. 

\subsection{An extension of Dirac's Large Number Hypothesis (LNH): the emergence of a minimum speed as a new constant of nature related to gravity and the cosmological constant}

We will also show the relationship between the minimum speed $V$ and the cosmological constant
$\Lambda$, so that the so-called ultra-referential (preferred referential) 
$S_{V}$ (Fig.2) associated with $V$ represents a kind of Machian background field (a vacuum energy) that leads to the cosmological constant $\Lambda$. In order to do that, we will start from Dirac's LNH by introducing the ratio of the forces of gravitational and electric interaction between an electron and a proton, namely:

 \begin{equation}
 \frac{F_e}{F_g}=\frac{e^2}{Gm_pm_e}=\sqrt{N}\sim 10^{40},  
 \end{equation}
 where $e^2=q_e^2/4\pi\epsilon_0$. $N(\sim 10^{80}$) is the well-known magic number of 
 Eddginton, which is of the order of the number of particles in the universe. $m_e$ and 
 $m_p$ are the electron and proton masses respectively. 
 
 The large number of the order of $10^{40}$ is the well-known Dirac's large number. Is is interesting to notice that such large number can also be obtained in other ways, as for instance, the ratio $F_e/F_g\sim r_p/R_H\sim 10^{40}$, where $r_p$ is the proton radius and $R_H$ is the Hubble radius. This indicates that such large number also connects length scales of the  micro-world (proton radius) with the macro-world (universe radius given by the Hubble radius). 

 We know that the orbital speed of the electron in the ground state of the Bohr's hydrogen atom is given as follows:

\begin{equation}
 v_B=v_{Bohr}=\alpha c =\frac{e^2}{\hbar}=\frac{q^2_e}{4\pi\epsilon_0\hbar}, 
\end{equation}
where $\alpha(=e^2/\hbar c=q^2_e/4\pi\epsilon_0\hbar c\approx 1/137$) is the fine structure 
constant of Coulombian origin and $v_B=v_{Bohr}(\approx (1/137) c\sim10^5m/s)$ is the velocity of the electron in the fundamental state of the hydrogen atom so-called Bohr velocity. 

Now by making an extension of Dirac's LNH, so that we use the work-energy theorem
to implement both works to ionize the real hydrogen atom (with Coulombian interaction) and a hypothetical hydrogen atom with only gravitational interaction between the proton and the electron to be carried from fundamental (Bohr) radius $a_0$ to infinite, we find the following ratios of work (of applied forces) and kinetic energy, namely: 

\begin{equation}
 \frac{-W_{F_e}(a_0\rightarrow\infty)}{-W_{F_g}(a_0\rightarrow\infty)}=
 \frac{W_{F_e}(\infty\rightarrow a_0)}{W_{F_g}(\infty\rightarrow a_0)}=
 \frac{\frac{q_e^2}{4\pi\epsilon_0}\int_{\infty}^{a_0}\frac{1}{r^2}dr}
 {Gm_pm_e\int_{\infty}^{a_0}\frac{1}{r^2}dr}\equiv\frac{F_e}{F_g}
 =\frac{q_e^2}{4\pi\epsilon_0Gm_pm_e}=
 \frac{\frac{1}{2}m_ev_B^2}{\frac{1}{2}m_ev_G^2}\sim 10^{40}, 
\end{equation}
where $v_G$ is the most fundamental velocity (a too small kinetic energy), since it has origin from the work of the gravitational force as being the negative of the same applied force to ionize a hypothetical gravitational hydrogen atom. 

From Eq.(3), we get

\begin{equation}
\frac{v_B^2}{v_G^2}=\frac{q_e^2}{4\pi\epsilon_0Gm_pm_e}\sim 10^{40},
\end{equation}
where $v_B$ is the Bohr velocity [Eq.(2)] 

By substituting Eq.(2) ($v_B$) in Eq.(4) of the extended LNH for $v_G$, and after by performing
the calculations, we finally find 

\begin{equation} 
v_G=\sqrt{\frac{Gm_em_p}{4\pi\epsilon_0}}\frac{q_e}{\hbar}, 
\end{equation}
where $v_G\approx 4.58\times 10^{-14}m/s$. 

Eq.(5) shows clearly the fundamental (lowest) speed $v_G$ due to its gravitational origin, since it depends on the constant of gravity $G$, i.e., $v_G\propto G^{1/2}$, such that, if 
gravity vanishes ($G\rightarrow 0$), $v_G$ would be zero; however, as the presence of gravity 
cannot be eliminated in anywhere, rest is prevented due to a zero-point energy associated
to the most fundamental vacuum energy of gravitational origin. In this sense, we are led to
postulate $v_G$ as being a new kinematic constant connected to such a vacuum of quantum gravity at very low energies, i.e., it is a unattainable minimum speed associated with a preferred reference frame of background field, which is also unattainable for any particles, so that the speed of any particles must belong to the interval $v_G<v\leq c$ within a new scenario of
Deformed Special Relativity (DSR) so-called Symmetrical Special Relativity (SSR), $v_G$ being the inferior cut-off of speed related to the vacuum energy. 

Let us simply use the notation $V$ for representing the universal minimum speed $v_G$, such that we write

\begin{equation}
\frac{F_e}{F_g}=\left(\frac{v_B}{v_G}\right)^2=\left(\frac{v_B}{V}\right)^2\sim 10^{40},
\end{equation}
where $V=v_G$ [Eq.(5)] represents the invariant kinematic aspect at lower energies in SSR.

It has also been shown that the existence of $V$ in the spacetime of SSR leads to the uncertainty principle\cite{uncertainty}.  

As we have already been able to obtain the universal minimum speed $V$ [Eq.(5)] by
means of the extended LNH given in Eq.(4) [or Eq.(6)], we will look for the relationship between $V$ and the cosmological constant $\Lambda$. To do this, we should first remember that the ratio of the Hubble radius ($R_H$) and the radius of the proton ($r_p$) is exactly of the order of the square root of the magic number of Eddginton ($\sqrt N\sim 10^{40}$)
with $N\sim 10^{80}$. So, let us just write Dirac's LNH, as follows:

\begin{equation}
\frac{F_e}{F_g}=\left(\frac{v_B}{V}\right)^2\sim\frac{R_H}{r_p}\sim 10^{40}, 
\end{equation}
where the proton radius $r_p\sim 10^{-15}$m coincides in being the classical electron radius, which is obtained to explain the energy of the electron ($m_ec^2$) as originating from the
Columbian self-interaction of the electron charge, i.e., $m_e c^2=q_e^2/4\pi\epsilon_0 r_{electron}$, from where one obtains $r_{electron}=r_{classical}\sim r_{proton}=r_p$. 

As we have $r_p\sim q_e^2/4\pi\epsilon_0 m_ec^2$, so by substituting $r_p$ in Eq.(7), we obtain 

\begin{equation}
\left(\frac{v_B}{V}\right)^2\sim\frac{4\pi\epsilon_0 m_ec^2R_H}{q_e^2}\sim 10^{40}. 
\end{equation}

In a previous work, where SSR theory was introduced\cite{3}, it has already been
shown that the cosmological constant depends on the Hubble radius ($R_H\sim 10^{26}m$), i.e., $\Lambda=6c^2/R_H^2\sim 10^{-35}s^{-2}$\cite{3}, where we obtain 

\begin{equation}
R_H=c\sqrt{\frac{6}{\Lambda}}. 
\end{equation}

We know that $v_B=q_e^2/4\pi\epsilon_0\hbar$. So by substituting Eq.(9) above in Eq.(8) and 
performing the calculations, we find 

\begin{equation}
v_G=V\sim\frac{q_e^3}{6^{1/4}(4\pi\epsilon_0 c)^{3/2} m_e^{1/2}\hbar}\Lambda^{1/4}.
\end{equation}

Thus we realize that the most fundamental state of vacuum associated with the minimum speed 
$V$ has direct connection to the cosmological constant $\Lambda$ (vacuum energy), whose 
equation of state (EOS) is $p(presuure)=-\rho(energy~density~of~vacuum)$, thus leading to 
the anti-gravity.

\subsection{The vacuum energy: the minimum speed and the cosmological conatant}

We realize that the EOS $p=-\rho$ is the limiting case of EOS associated with the cosmological vacuum (cosmological constant $\Lambda$) connected to the minimum speed, i.e., 
$V\propto\Lambda^{1/4}$ [Eq.(10)].

SSR will be able to describe in detail a superfluid, which is very similar to what we see in the literature on Gravastar/Dark Energy Star.

Our goal is to investigate a more complex structure of GBEC of gravastar, which has many vibrational modes of vacuum, so that the single mode of EOS ($p=-\rho$) for a given cosmological constant $\Lambda$ should be generalized for a variable cosmological parameter ($\Lambda(r)$) inside the spherical repulsive core of GBEC with radius $r$, where $r\leq R_{core}$ (section 4). 

\subsection{Spacetime transformations with an invariant minimum speed}

A breakdown of Lorentz symmetry for very low energies\cite{3}\cite{4}\cite{5}\cite{6} generated
by the presence of a background field (a vacuum energy related to the cosmological constant)
creates a new causal structure in spacetime, where we have a mimimum speed $V$, which is 
unattainable for all particles, and also a universal dimensionless constant $\xi$\cite{3}, which couples the gravitational field to the electromagnetic one ($c$), as shown in a previous work\cite{3}, namely:

\begin{equation}
\xi=\frac{v_G}{c}=\frac{V}{c}=\sqrt{\frac{Gm_{p}m_{e}}{4\pi}}\frac{q_{e}}{\hbar c},
\end{equation}
where $m_{p}$ and $m_{e}$ are respectively the mass of the proton and electron. The value of such a minimum speed is $v_G=V=4.5876\times 10^{-14}$ m/s. 
We find $\xi=1.5302\times 10^{-22}$\cite{3}. 

Therefore the light cone contains a new region of causality called {\it dark cone}\cite{3}, so that the velocities of the particles must belong to the following range: $V$(dark cone)$<v<c$ (light cone) (Fig.1). 

The breaking of Lorentz symmetry group destroys the properties of the transformations of Special Relativity (SR) and so generates intriguing kinematics and dynamics for speeds very close to the minimum speed $V$, i.e., for $v\rightarrow V$, we find new relativistic effects such as the contraction of the improper time and the dilation of space\cite{3}\cite{experiment}. In this new scenario, the proper time also suffers relativistic effects such as its own dilation with regard to the improper one when $v\rightarrow V$\cite{3}\cite{4}, namely:

\begin{equation}
\Delta\tau\sqrt{1-\frac{V^{2}}{v^{2}}}=\Delta t\sqrt{1-\frac{v^{2}}{c^{2}}}.
\end{equation}

As the minimum speed $V$ is an invariant quantity as the speed of light $c$, $V$ does not alter the value of the speed $v$ of any particle\cite{3}\cite{4}. Therefore we have called ultra-referential $S_{V}$\cite{3}\cite{4} as being the preferred (background) reference frame in
relation to which we have the speeds $v$ of any particle (Fig.2). In view of this, the reference frame transformations change substantially in the presence of $S_V$, as shown first in the special case $(1+1)D$, namely:

\begin{figure}
\begin{center}
\includegraphics[scale=0.90]{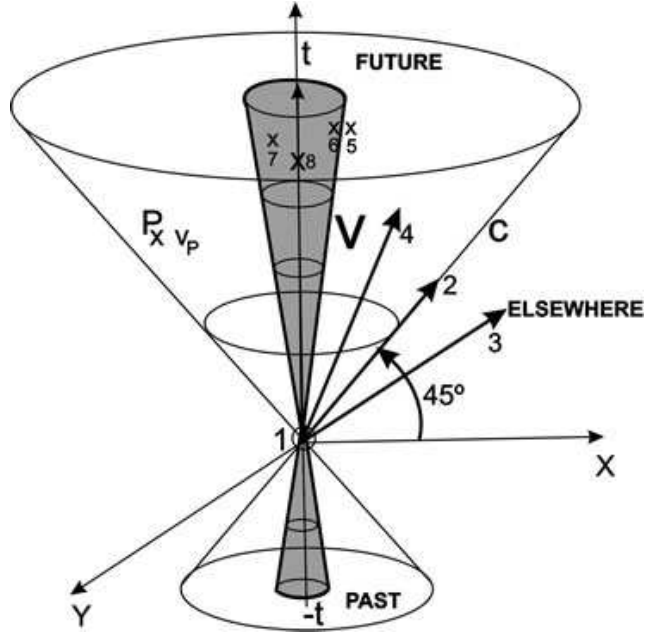}
\end{center}
\caption{The external and internal conical surfaces represent respectively the speed of light 
$c$ and the unattainable minimum
speed $V$, which is a definitely prohibited boundary for any particle. For a point $P$ in the world line of a particle, in the interior
of the two conical surfaces, we obtain a corresponding internal conical surface, such that we must have $V<v_p\leq c$. The $4$-interval
$S_4$ is of type time-like. The $4$-interval $S_2$ is a light-like interval (surface of the light cone). 
The $4$-interval $S_3$ is of type space-like (elsewhere). The novelty in spacetime of SSR are the $4$-intervals $S_5$ (surface of
the dark cone) representing an infinitly dilated time-like interval\cite{3}, including the $4$-intervals $S_6$, $S_7$ and 
$S_8$ inside the dark cone for representing a new space-like region\cite{3}.}
\end{figure}

\begin{figure}
\begin{center}
\includegraphics[scale=0.90]{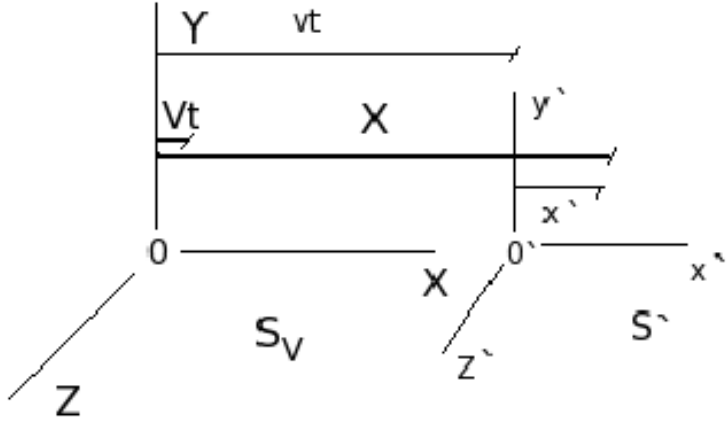}
\end{center}
\caption{The reference frame $S^{\prime}$ moves in $x$-direction with a speed $v(>V)$ with respect to the universal background field connected to the unattainable (absolute) ultra-referential $S_V$ associated with $v_G=V$. If $v_G=V\rightarrow 0$ (or even $G=0$ in Eq.(5)), $S_V$ is eliminated (no vacuum energy of quantum gravity) and, thus the galilean (inertial)
frame $S$ takes place, recovering the Lorentz transformations.}
\end{figure}

\begin{equation}
x^{\prime}=\Psi(X-vt+v_Gt)=\Psi(X-vt+Vt)
\end{equation}

and 

\begin{equation}
 t^{\prime}=\Psi(t-vX/c^2+v_GX/c^2)=\Psi(t-vX/c^2+VX/c^2), 
 \end{equation}
where $v_G=V=\sqrt{Gm_pm_e}e/\hbar$ and $\Psi=\sqrt{1-v_G^2/v^2}/\sqrt{1-v^2/c^2}=
\sqrt{1-V^2/v^2}/\sqrt{1-v^2/c^2}$. 

As the transformations above in Eq.(13) and Eq.(14) are given in $(1+1)D$ for the simple case of one dimensional motion as an approximation of the real motion given in $(3+1)D$ spacetime of SSR, they just appear in their scalar form, so that we simply consider the scalar $V$ for representing the minimum speed at one spatial dimension, which just represents an ideal case. However, it is important to stress that the real case of $(3+1)D$ spacetime has a $3D$-vectorial background field represented by the vector $\vec V$ that breaks the Lorentz symmetry, being invariant at any direction of $3D$-space (Fig.3), in such a way that it reduces to the scalar
$V$ at the ideal case of one spatial dimension. The general case $(3+1)D$ with the background $3$-vector $\vec V$ (Fig.3) will be shown soon. This will ensure that rest must be prevented in the real case of $(3+1)D$ spacetime of SSR. 

The transformations shown in Eq.(13) and Eq.(14) are the direct transformations
from $S_V$ [$X^{\mu}=(ct,X)$] to $S^{\prime}$ [$x^{\prime\nu}=(ct^{\prime},x^{\prime})$], where
we have $x^{\prime\nu}=\Lambda^{\nu}_{\mu} X^{\mu}$ ($x^{\prime}=\Lambda X$),
so that we write the matrix of transformation for one-dimensional motion in 
$x$-direction (Fig.2), as follows: 

\begin{equation}
\displaystyle\Lambda= 
\begin{pmatrix}
\theta\gamma & -\theta\gamma\beta_* & 0 & 0\\
-\theta\gamma\beta_* & \theta\gamma & 0 &  0\\
  0 & 0 & 1 & 0\\
 0  & 0 & 0 & 1
\end{pmatrix},
\end{equation}
or simply

\begin{equation}
\displaystyle\Lambda=
\begin{pmatrix}
\Psi & -\Psi\beta_* \\
-\Psi\beta_* & \Psi
\end{pmatrix},
\end{equation}
such that we recover $\Lambda\rightarrow\ L$ (Lorentz matrix of rotation) for
$\alpha\rightarrow 0$, which implies $\Psi\rightarrow\gamma$ of SR. 

We have $\Psi=\theta\gamma$ and $\beta_{x*}=\beta_*=\beta(1-\alpha)$, as $v=v_x$ for one-dimensional motion in $x$-direction (Fig.2). 

We obtain $det\Lambda=\frac{(1-\alpha^2)}{(1-\beta^2)}[1-\beta^2(1-\alpha)^2]$, 
where $0<det\Lambda<1$. Since $V$ ($S_V$) is unattainable ($v>V)$, this assures that 
$\alpha=V/v<1$ and therefore the matrix $\Lambda$ admits inverse ($det\Lambda\neq 0$ $(>0)$). However, $\Lambda$ is a non-orthogonal matrix ($det\Lambda\neq\pm 1$) and so it does not represent a rotation matrix ($det\Lambda\neq 1$) in SSR\cite{3}

Actually the result $det\Lambda\approx 0$ for $\alpha\approx 1$ or $v\approx V$ emerges from a new relativistic effect of SSR for treating very low energies at a ultra-infrared regime
(very large wavelengths) too close to the background frame $S_V$, i.e., $v\approx V$. 

The inverse transformations (from $S^{\prime}$ to $S_V$) are

 \begin{equation}
 X=\Psi^{\prime}(x^{\prime}+\beta_{*}ct^{\prime})=\Psi^{\prime}(x^{\prime}+vt^{\prime}-Vt^{\prime}),
\end{equation}

 \begin{equation}
 t=\Psi^{\prime}\left(t^{\prime}+\frac{\beta_{*}
 x^{\prime}}{c}\right)=\Psi^{\prime}\left(t^{\prime}+\frac{vx^{\prime}}{c^2}-
\frac{Vx^{\prime}}{c^2}\right).
  \end{equation}

In matrix form, we get the inverse transformation $X^{\mu}=\Lambda^{\mu}_{\nu} x^{\prime\nu}$ ($X=\Lambda^{-1}x^{\prime}$), so that the inverse matrix is

\begin{equation}
\displaystyle\Lambda^{-1}=
\begin{pmatrix}
\Psi^{\prime} & \Psi^{\prime}\beta_* \\
\Psi^{\prime}\beta_* & \Psi^{\prime}
\end{pmatrix},
\end{equation}
where we can show that $\Psi^{\prime}$=$\Psi^{-1}/[1-\beta^2(1-\alpha)^2]=\theta^{-1}\gamma^{-1}/(1-\beta_*^2)$, so that we must satisfy $\Lambda^{-1}\Lambda=I$.

 Indeed we have $\Psi^{\prime}\neq\Psi$ and therefore $\Lambda^{-1}\neq\Lambda(-v)$. This aspect of $\Lambda$ has an important physical implication. In order to understand this implication, let us first consider the rotation aspect of Lorentz matrix in SR. Under SR, we
 have $\alpha=0$ ($V=0$), so that $\Psi^{\prime}\rightarrow\gamma^{\prime}=\gamma=(1-\beta^2)^{-1/2}$. This symmetry ($\gamma^{\prime}=\gamma$, $L^{-1}=L(-v)$) happens because the Galilean reference frames permit to exchange the speed $v$ (of $S^{\prime}$) for $-v$ (of $S$) when we are at rest at $S^{\prime}$. However, in SSR, as there is no rest at $S^{\prime}$, we cannot change $v$ (of $S^{\prime}$) for $-v$ (of $S_V$) due to that asymmetry ($\Psi^{\prime}\neq\Psi$, $\Lambda^{-1}\neq\Lambda(-v)$), thus leading to Lorentz violation. Due to this fact,
 $S_V$ must be covariant, namely $V$ remains invariant for any change of reference frames in such spacetime. This issue will be well-understood for the general case $(3+1)D$, where 
 we have an isotropic $3$-vectorial background field $\vec V$, thus preventing rest
 ($v=0$) for any particles. 
 
\begin{figure}
\begin{center}
\includegraphics[scale=1.0]{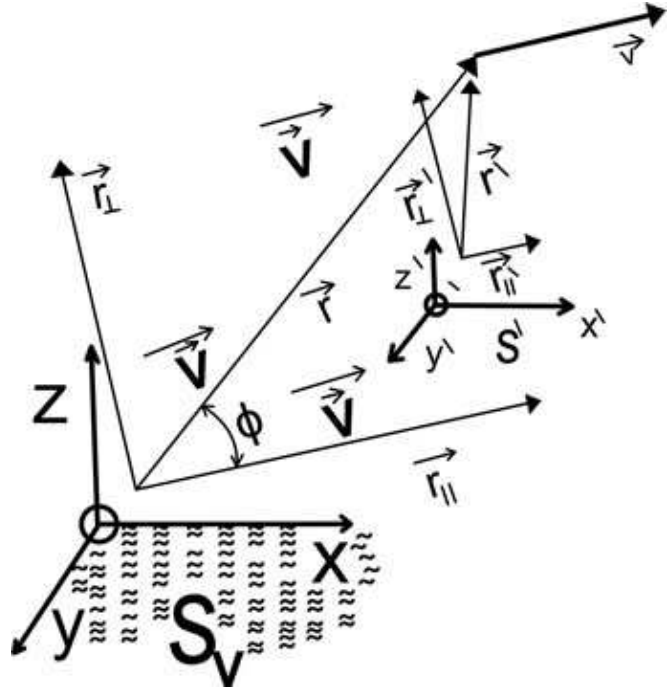}
\end{center}
\caption{$S^{\prime}$ moves with a $3D$-velocity $\vec v=(v_x,v_y,v_z)$ in relation to $S_V$. For the special case of $1D$-velocity
$\vec v=(v_x)$, we recover the case $(1+1)D$; however, in this general case of $3D$-velocity $\vec v$, there must be a background vector
 $\vec V$ (minimum velocity)\cite{3} with the same direction of $\vec v$ as shown in this figure. Such a background vector $\vec V=(V/v)\vec v$ is
related to the background reference frame (ultra-referential) $S_V$, thus leading to a
Lorentz violation. The modulus of $\vec V$ is invariant at any direction.}
\end{figure}

The $(3+1)D$ (Fig.3)\cite{3} transformations in SSR are

\begin{equation}
\vec r^{\prime}=\theta\left[\vec r + (\gamma-1)\frac{(\vec r.\vec v)}{v^2}\vec v-\gamma\vec v(1-\alpha)t\right]=
\theta\left[\vec r + (\gamma-1)\frac{(\vec r.\vec v)}{v^2}\vec v-\gamma\vec vt+\gamma\vec Vt\right],
\end{equation}
where $\theta=\sqrt{1-\frac{V^{2}}{v^{2}}}$ and
$\theta\gamma=\Psi=\frac{\sqrt{1-\frac{V^{2}}{v^{2}}}}{\sqrt{1-\frac{v^{2}}{c^{2}}}}$.

And

\begin{equation}
t^{\prime}=\theta\gamma\left[t-\frac{\vec r.\vec v}{c^2}(1-\alpha)\right]=
\theta\gamma\left[t-\frac{\vec r.\vec v}{c^2}+\frac{\vec r.\vec V}{c^2}\right],
\end{equation}
where $\vec V=(\vec v/v)V$. 
It is easy to verify that, if we have $\vec v||\vec r(\equiv X\vec e_x)$, we recover Eq.(13) for $(1+1)D$ spacetime. So, in the special case $(1+1)D$ with $\vec v=v_x=v$, we find the following transformations: $x^{\prime}=\Psi(x-vt+Vt)$ and $t^{\prime}=\Psi(t-vx/c^2+Vx/c^2)$. The inverse transformations for the general case $(3+1)D$ and $(1+1)D$ were demonstrated in
ref.\cite{3}. Of course, if we make $V\rightarrow 0$, we recover the well-known Lorentz transformations.

Putting the transformations given in Eq.(20) and Eq.(21) into a matricial form, we obtain the following matrix:

\begin{equation}
\displaystyle\Lambda_{(4X4)}=
\begin{pmatrix}
\theta\gamma  &-\theta\gamma\beta_{x*}  &-\theta\gamma\beta_{y*}  &-\theta\gamma\beta_{z*} \\
 -\theta\gamma\beta_{x*} & \left[\theta+\theta(\gamma-1)\frac{\beta_x^2}{\beta^2}\right]  & \left[\theta(\gamma-1)\frac{\beta_x\beta_y}{\beta^2}\right] & \left[\theta(\gamma-1)\frac{\beta_x\beta_z}{\beta^2}\right]\\
 -\theta\gamma\beta_{y*}                       &\left[\theta(\gamma-1)\frac{\beta_y\beta_x}{\beta^2}\right]  & \left[\theta+\theta(\gamma-1)\frac{\beta_y^2}{\beta^2}\right]  & \left[\theta(\gamma-1)\frac{\beta_y\beta_z}{\beta^2}\right] \\
 -\theta\gamma\beta_{z*}                         & \left[\theta(\gamma-1)\frac{\beta_z\beta_x}{\beta^2}\right]           & \left[\theta(\gamma-1)\frac{\beta_z\beta_y}{\beta^2}\right]    &  \left[\theta+\theta(\gamma-1)\frac{\beta_z^2}{\beta^2}\right]
\end{pmatrix}, 
\end{equation}
where we have defined the compact notations namely $\beta_{x*}=\beta_{x}(1-\alpha)$, $\beta_{y*}=\beta_{y}(1-\alpha)$ and $\beta_{z*}=\beta_{z}(1-\alpha)$. 

Writing the general matrix of transformation $\Lambda_{(4X4)}$ [Eq.(22)] in a compact form $(2\times 2)$, we find 

\begin{equation}
\displaystyle\Lambda_{(2\times 2)}=
\begin{pmatrix}
\theta\gamma & -\frac{\theta\gamma {\bf v}^T(1-\alpha)}{c} \\
-\frac{\theta\gamma{\bf v}(1-\alpha)}{c} & \left[\theta I+\theta(\gamma-1)\frac{{\bf v}{\bf v^T}}{v^2}\right]
\end{pmatrix},
\end{equation}
where $I=I_{3\times 3}$ is the identity matrix and ${\bf v}^T=(v_x, v_y, v_z)$ is the transposed of ${\bf v}$.

If we make $\alpha=0$ ($V=0$), which implies $\theta=1$, the matrix in Eq.(22) (or Eq.(23)) recovers the general Lorentz matrix.  

The $(3+1)D$ (Fig.3) inverse transformations in SSR were also obtained before\cite{3}, namely:

\begin{equation}
\vec r=\theta^{-1}\vec r^{\prime} + \theta^{-1}\left[\left(\frac{\gamma^{-1}}{1-\beta^2_*}-1\right)
\left(\frac{\vec r^{\prime}.\vec v}{v^2}\right)+
\frac{(\gamma^{-1})_*}{1-\beta^2_*}t^{\prime}\right]\vec v 
\end{equation}
where we have used the simplified notation $\beta_*=\beta(1-\alpha)$. We also 
have $(\gamma^{-1})_*=\gamma^{-1}(1-\alpha)$.

And 

\begin{equation}
t=\frac{\theta^{-1}\gamma^{-1}}{1-\beta^2(1-\alpha)^2}\left[t^{\prime}+ \frac{\vec r^{\prime}.\vec v}{c^2}(1-\alpha)\right]=
\frac{\theta^{-1}\gamma^{-1}}{1-\beta^2(1-\alpha)^2}\left[t^{\prime}+\frac{\vec r^{\prime}.\vec v}{c^2}-
\frac{\vec r^{\prime}.\vec V}{c^2}\right].
\end{equation}

In Eq.(24) and Eq.(25), if we make $\alpha=0$ (or $\vec V=0$), we recover the $(3+1)D$ Lorentz inverse transformations.

From Eq.(24) and Eq.(25), we obtain the general inverse matrix of transformation as follows:

\begin{equation}
\displaystyle\Lambda^{-1}_{(4\times 4)}=
\begin{pmatrix}
\frac{\theta^{-1}\gamma^{-1}}{1-\beta_*^2}  &\frac{\theta^{-1}\gamma^{-1}\beta_{x*}}{1-\beta_*^2}  &\frac{\theta^{-1}\gamma^{-1}\beta_{y*}}{1-\beta_*^2}  &\frac{\theta^{-1}\gamma^{-1}\beta_{z*}}{1-\beta_*^2} \\
 \frac{\theta^{-1}\gamma^{-1}\beta_{x*}}{1-\beta_*^2} & \left[\theta^{-1}+\theta^{-1}\left(\frac{\gamma^{-1}}{1-\beta_*^2}-1\right)\frac{\beta_x^2}{\beta^2}\right]  & \left[\theta^{-1}\left(\frac{\gamma^{-1}}{1-\beta_*^2}-1\right)\frac{\beta_x\beta_y}{\beta^2}\right] & \left[\theta^{-1}\left(\frac{\gamma^{-1}}{1-\beta_*^2}-1\right)\frac{\beta_x\beta_z}{\beta^2}\right]\\
 \frac{\theta^{-1}\gamma^{-1}\beta_{y*}}{1-\beta_*^2}                       &\left[\theta^{-1}\left(\frac{\gamma^{-1}}{1-\beta_*^2}-1\right)\frac{\beta_y\beta_x}{\beta^2}\right]  & \left[\theta^{-1}+\theta^{-1}\left(\frac{\gamma^{-1}}{1-\beta_*^2}-1\right)\frac{\beta_y^2}{\beta^2}\right]  & \left[\theta^{-1}\left(\frac{\gamma^{-1}}{1-\beta_*^2}-1\right)\frac{\beta_y\beta_z}{\beta^2}\right] \\
 \frac{\theta^{-1}\gamma^{-1}\beta_{z*}}{1-\beta_*^2}                      & \left[\theta^{-1}\left(\frac{\gamma^{-1}}{1-\beta_*^2}-1\right)\frac{\beta_z\beta_x}{\beta^2}\right]           & \left[\theta^{-1}\left(\frac{\gamma^{-1}}{1-\beta_*^2}-1\right)\frac{\beta_z\beta_y}{\beta^2}\right]    &  \left[\theta^{-1}+\theta^{-1}\left(\frac{\gamma^{-1}}{1-\beta_*^2}-1\right)\frac{\beta_z^2}{\beta^2}\right]
\end{pmatrix}, 
\end{equation}
where we have $\beta_{x*}=\beta_{x}(1-\alpha)$, $\beta_{y*}=\beta_{y}(1-\alpha)$, $\beta_{z*}=\beta_{z}(1-\alpha)$
and $\beta_{*}=\beta(1-\alpha)=v(1-\alpha)/c=v_{*}/c$.

Writing the general inverse matrix of transformation $\Lambda^{-1}$ [Eq.(26)] in a compact form $(2\times 2)$, we have

\begin{equation}
\displaystyle\Lambda^{-1}_{(2\times 2)}=
\begin{pmatrix}
\frac{\theta^{-1}\gamma^{-1}}{1-\beta^2_*} & \frac{\theta^{-1}\gamma^{-1}{\bf v}^T_*}{c(1-\beta^2_*)} \\
\frac{\theta^{-1}\gamma^{-1}{\bf v_*}}{c(1-\beta^2_*)} &
\left[\theta^{-1}I+\theta^{-1}(\frac{\gamma^{-1}}{1-\beta^2_*}-1)\frac{{\bf v}{\bf v^T}}{v^2}\right]
\end{pmatrix},
\end{equation}
where ${\bf v}^T_*={\bf v}^T(1-\alpha)$, ${\bf v}_*={\bf v}(1-\alpha)$ and $\beta_*=\beta(1-\alpha)$.

 We can compare the inverse matrix in Eq.(27) with the matrix in Eq.(23) and verify that 
 $\Lambda^{-1}\neq\Lambda^T$, in a similar way as made before for the particular case $(1+1)D$ (one dimensional motion). 

If we make $\alpha=0$ ($\vec V=0$), which implies $\theta=1$, the inverse matrix in Eq.(26) (or Eq.(27)) recovers the general inverse matrix of Lorentz. 

Although we associate the minimum speed $V$ with the ultra-referential $S_{V}$, this is inaccessible for any particle. Thus, the effect
of such new causal structure of spacetime (SSR) generates a symmetric mass-energy effect to what happens close to the speed of light $c$, i.e., it was shown
that $E=m_0c^2\Psi$, so that $E\rightarrow 0$ when $v\rightarrow V$\cite{3}\cite{4}. It was also shown that the minimum speed $V$
is associated with the cosmological constant, which is equivalent to a fluid (vacuum energy) with negative pressure\cite{3}\cite{4}: just reminding that we have shown that 
$V\propto\Lambda^{1/4}$ in Eq.(10).

The metric of such symmetrical spacetime of SSR is a deformed Minkowsky metric with a global multiplicative function $\Theta$, i.e., a conformal factor $\Theta$\cite{Rodrigo}, being 
similar to a dS-metric, namely:

\begin{equation}
dS^{2}={\Theta}g_{\mu\nu}dx^{\mu}dx^{\nu},
\end{equation}
where $\Theta=\theta^{-2}=1/(1-V^2/v^2)$\cite{3}\cite{4}\cite{Rodrigo}.

We can say that SSR geometrizes the quantum phenomena as investigated before (the Uncertainty Principle)\cite{5}\cite{uncertainty} in order to allow us 
to associate quantities belonging to the microscopic world with a new geometric structure that originates from Lorentz symmetry breaking. Thus SSR may be a candidate to try to solve the problems associated with the gravitational collapse, which is a phenomenon that mixes 
inevitably quantum properties with the geometric structure of spacetime.

\section{The GBEC and its anisotropy}

The core of a Gravastar/Dark Energy Star is described as being composed of an exotic material called Gravitational Bose Einstein Condensate (GBEC)\cite{9}\cite{16}\cite{17}. This is a relativistic superfluid and this region connects with a shell of ordinary matter (baryonic matter) described by the Schwarzschild metric (Fig.4). Such a connection would
take place by means of a phase transition in spacetime\cite{12}\cite{14}\cite{15} that occurs near the Schwarzschild radius. Thus, by following the works of CFV and MM\cite{9}\cite{10}\cite{13}\cite{16}\cite{17}\cite{24}, we write the metric of a gravastar\cite{16}\cite{17}, namely:

\begin{equation}
dS^{2}=-f(r)c^{2}dt^{2}+\frac{dr^{2}}{h(r)}+r^2d\Omega^{2},
\end{equation}
where $d\Omega$ is the well-known solid angle. The metric functions $f(r)$ and $h(r)$ are given for dS-sector (GBEC), namely:

\begin{equation}
f_{GBEC}(r)=A\left(1-\frac{r^2}{R^2_{core}}\right)
\end{equation}
and
\begin{equation}
h_{GBEC}(r)=\left(1-\frac{r^{2}}{R^2_{core}}\right), 
\end{equation}
where $R_c=R_{core}$ (core radius) and $A$ is a given constant or even a certain function, which is obtained depending on the boundary conditions. In our investigation, as we will see there is a similarity between GBEC-metric (a kind of dS-metric) and SSR-metric, we will make a mapping between them in the next section. 

The constant vacuum energy density $\rho$ inside the simple model of GBEC with 
a single positive cosmological constant $\Lambda$ (dS-space) as shown in Fig.4 is 

\begin{equation}
\rho=\frac{\Lambda c^2}{8\pi G},
\end{equation}
where $\Lambda$ is the cosmological constant whose the unique vacuum state is 
represented by the well-known EOS of vacuum energy, namely: 

\begin{equation}
 p=w\rho=-\rho,
\end{equation}
where $w=-1$, $p$ is the pressure and $\rho$ is the vacuum energy density.

On the other hand, the baryonic region with ultra-relativistic plasma in shell (Fig.4) is described by the following EOS: 

\begin{equation}
p=w\rho=\rho,
\end{equation}
with $w=+1$ by representing the attractive matter of the ultra-relativistic plasma in shell. 

Here it is important to call attention to the fact that we are proposing a general model of gravastar, i.e., a dark energy star so that we have a more general EOS that encompasses several vibrational degrees of vacuum in order to explain the anisotropy inside GBEC, in such a way that
the pressure becomes practically zero at the core radius, which is equivalent to $v\approx V$
in a general EOS, so that $\frac{dp}{d\rho}=w(v)=-\Omega^{2}=-\frac{v^2}{c^2}$,
where $\Omega=\beta=v/c$\cite{11}\cite{12}. Thus we obtain the following general EOS inside 
GBEC of our model of dark energy star, namely:
 
\begin{equation}
p=-w(v)\rho=-\frac{v^2}{c^2}\rho. 
\end{equation}

In the general EOS given by Eq.(35), we can see that there is a correspondence of the 
well-known EOS for the fundamental vacuum energy of $\Lambda$, i.e., $p=-\rho$ with 
$v=c$, which represents the maximum repulsive pressure $p$ inside GBEC at $r=0$ or 
exactly the center of the spherical repulsive core of the dark energy star, where the 
so-called anisotropy is $\Delta=0$ (subsection A). In other words, we can say that we find 
$p=-\rho$ associated to the maximum repulsive parameter $\Lambda(r=0)=\Lambda_0$ in the center of the core. But, when we approaches to $r=R_{core}$, Eq.(35) shows us that the repulsive pressure is practically null at the surface of the core, i.e., 
$p=-(V^2/c^2)\rho=-\xi^2\rho\approx 0$ with $v=V$. This important result already indicates firstly that there is a direct relationship between $v=V$ and $r=R_{core}$ in the sense 
that both $r=R_{core}$ and $v=V$ lead to the divergenges of GBEC-metric [Eq.(29)] and 
SSR-metric in Eq.(28) (a kind of dS-(conformal) metric\cite{Rodrigo}) respectively. Therefore
we are led to think that there is a direct mapping between Eq.(28) and Eq.(29), where the coefficient $A$ [Eq.(30)] should be adjusted for consistency between both metrics whose boundary conditions ($V$ and $R_{core}$) are equivalent. This issue will be treated in the next section
(section 4). 

Furthermore, the decreasing of the repulsive pressure $p$ to quasi-zero [Eq.(35)] when the radius $r$ approaches to $R_{core}$, so that the parameter $\Lambda(r=R_{core})\approx 0$, allows us to understand the emergence of the phase transition from anti-gravity to gravity as there should be a coexistence region close to $R_{core}$, i.e., $R_{coexistence}<R_{core}$, where GBEC-metric [Eq.(29)] does not diverge with $p=0$ ($\Lambda_{coexistence}=0$). 

Therefore, the emergence of such null pressure at $r=R_{coexistence}$ is the unique way to permit the change of its signal to positive ($p>0$) when $r>R_{coexistence}$, where the attractive matter (baryonic plasma in shell) prevails as being gravity represented by the Schwarzchild metric. Actually, here it should be emphasized that such change of a negative
pressure to a positive pressure at $R_{coexistence}$ with $p=0$ is due to the variation of 
anisotropy $\Delta$ inside GBEC as we will investigate in the subsection A. The variation 
of the anisotropy $\Delta(r)$ is consistent with the general EOS given in Eq.(35), but 
the existence of anisotropy is not consistent with the single EOS $p=-\rho$ for a constant
cosmological constant $\Lambda>0$ inside the simple model of gravastar (Fig.4), as the single EOS leads to the absence of anisotropy ($\Delta=0$), which is valid only in the center of the 
repulsive core ($r=0$). So, if GBEC were really governed only by the EOS $p=-\rho$, the phase 
transition would not occur. This is the failure of the usual model of gravastar (Fig.4). 

Fig.6 shows the region of phase transition ($R_{core}$) with a certain approximation, in spite of this figure given in the literature\cite{9}\cite{10} is not able to show clearly the small
difference between $R_{coexistence}$ where $p$ must be exactly zero, and $R_{core}$ at which 
Cattoen\cite{9} consider that the pressure vanishes, since Cattoen does not make a clear 
distinction between $R_{core}$ and $R_{coexistence}$. However, thanks to the unattainable minimum speed $V$ of SSR, the little difference between both radius will be elucidated in the 
section 5, thus preventing the divergence of GBEC and Schwarzchild metric at $R_{coexistence}$,
so that the singularily of event horizon is eliminated. 

\begin{figure}
\begin{center}
\includegraphics[scale=0.3]{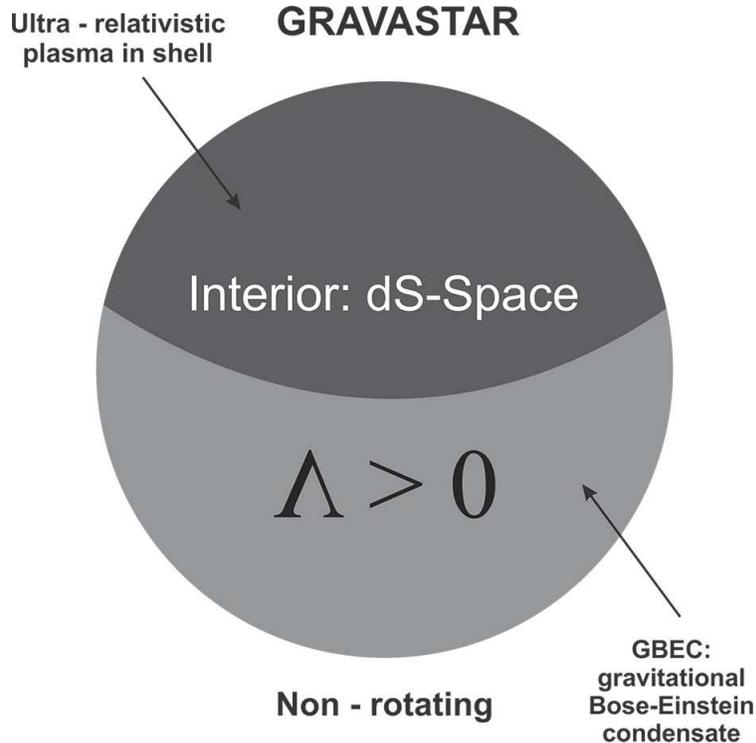}
\end{center}
\caption{In the interior of a usual model of gravastar, we find a repulsive core (GBEC) associated with a single positive cosmological constant $\Lambda>0$ (anti-gravity)\cite{9}. In this simple (ideal) model, GBEC is represented by its single cosmological constant and it is covered with a thin baryonic shell described by the Schwarzschild metric. But, in the present
model of SSR for dark energy star, GBEB is now described by a more general EOS given by 
Eq.(35), which is consistent with the anisotropy $\Delta$ (subsection A of section 3) since it leads to a null pressure in the region of coexistence phases. So it must be stressed that the 
single EOS $p=-\rho$ ($w=-1$) for a unique $\Lambda>0$ inside GBEC is not able to explain the 
anisotropy that leads to such null pressure at $R_{coexistence}$ and thus the phase transition.}
\end{figure}

\subsection{The relationship of DEC and WEC with the anisotropy and compactness}

The study of fluid with negative pressure in stars was regulated in refs.\cite{9}\cite{10}\cite{11}\cite{13} by following the Buchdahl-Bondi relation. We have the following conditions:

1) The NEC (The Null Energy Condition)

\begin{equation}
 \rho+p_{i}\geq 0.
\end{equation}

2) The DEC (The Dominant Energy Condition)

\begin{equation}
 |{p_{i}}|\leq\rho.
\end{equation}

The imposition of such conditions\cite{9}\cite{10}\cite{11}\cite{13} for the compactness of the star with mass $m$ results in a range of values that the compactness should obey in such a way that the horizon event is not formed. Thus, for the existence of a phase transition with the appearance of a repulsive core (GBEC), the values of compactness must conform to the following ranges in the respective layers close to the Schwazschild radius defined by the value
CFV\cite{9}, by respecting NEC and DEC conditions of Buchdahl-Bondi, namely:

\begin{equation}
\frac{8}{9}<\frac{2Gm}{c^{2}r}<1, ~~ \Delta\geq\frac{1}{4}\frac{\frac{2Gm}{c^{2}r}}{1-\frac{2Gm}{c^{2}r}}>0,
\end{equation}
where $\Delta$ is the magnitude of the so-called anisotropy, namely 
$\Delta=\frac{p_{t}-p_{r}}{\rho}$, where $p_t$ is the tangencial pressure and $p_r$ is the radial pressure inside the repulsive core with radius $R_{core}$. 

The coupling between the compactness and anisotropy characterizes the need to prevent
the formation of the event horizon.

Usually, in the literature about nuclear Astrophysics, the anisotropy is related to the presence of an electromagnetic field, but here the situation is different since the anisotropic term is introduced, so that we can obtain a repulsive effect, which is capable of preventing the
formation of the event horizon. This is why the connection between anisotropy and compactness is essential, which means that the anisotropy arises to respect the values above, being in accordance with the compactness that must be $\frac{2Gm}{c^{2}r}<1$. This means that
$r>R_S=2Gm/c^2$, i.e., a radius which is slightly larger than $R_S$, by preventing the emergence
of the event horizon. 

\section{Mapping between dS-metric that governs GBEC and SSR metric}

In this section, we map the geometric (metric) structure of SSR [Eq.(28)]\cite{3}\cite{4}\cite{5}\cite{6} into the geometry (metric) of spacetime of GBEC [Eq.(29)], as there is a 
similarity between them. 

We already know the dS-metric\cite{16}\cite{17}. So let us now rewrite Eq.(29) with azimuthal symmetry as the repulsive core is here considered to be a perfect sphere, where we only have 
$\Lambda=\Lambda(r)$, which does not depend on the angles $\theta$ and $\phi$, so that 
we simply neglect the term of solid angle in the metric, namely:  

\begin{equation}
 dS^{2}=-f(r)c^{2}dt^{2}+\frac{dr^{2}}{h(r)},
\end{equation}
where we already know the dS-metric functions of GBEC, i.e., 

\begin{equation}
f_{dS}(r)=A\left(1-\frac{r^2}{R^{2}_{core}}\right),~ h_{dS}(r)=\left(1-\frac{r^2}{R^2_{core}}\right).
\end{equation} 

Now we should rewrite the metric of SSR\cite{3}\cite{4}\cite{5}\cite{6} by considering the
effect of the deformed light cone with $c^{\prime}<<c$\cite{18}\cite{25} and also the 
deformed dark cone with $V^{\prime}>>V$ inside the collapsed star, being close to the hypothetical event horizon, i.e., we find that both cones approach to each other close to
the coexistence region of phase transition: see Fig.5 for the classical case of a black hole, where there is only the deformed light cone. So there is no dark cone in Fig.5, i.e., there is
just a drastic decreasing of $c$ close to $R_S$, so that we find $c^{\prime}(<<c)\rightarrow 0$
at the singular radius $r=R_S$, which does not occur in the non-classical case, where the 
light cone cannot become completely closed ($c^{\prime}>0$) due to the internal (repulsive) dark cone (Fig.1) that prevents its closing, avoiding the event horizon, since $V^{\prime}>>V$. We
will investigate the behavior of both $c^{\prime}$ and $V^{\prime}$ in section 5. 

Let us now write a given SSR-metric for representing a spherical repulsive core (with azimuthal symmetry) inside which the dark cone opens ($V'>V$) when $r$ goes to $R_{core}$, namely: 

\begin{equation}
dS^{2}=-\frac{c^{2}dt^{2}}{\left(1-\frac{V'^{2}}{v^{2}}\right)}+\frac{dr^{2}}{\left(1-\frac{V'^{2}}{v^{2}}\right)}, 
\end{equation}
from where $V$ changes to $V'>V$ (for a hypothetical internal observer inside the core), but $c$ mantains fixed for him (her) as it just changes to $c'<c$ for another exernal observer out of the repulsive core given by matter during the collapse of the baryonic sector, from where $V$ already mantains fixed for such external observer. 

In sum, for an internal observer inside the repulsive core, $V'>V$ (the dark cone opens) and the light cone remains fixed ($c$), while for an external observer in the baryonic sector (matter), 
$c'<c$ (the light cone closes) and the dark cone remains fixed ($V$), as both observers cannot
see clearly beyond the region of phase transition (quasi event horizon). 

\begin{figure}
\begin{center}
\includegraphics[scale=0.40]{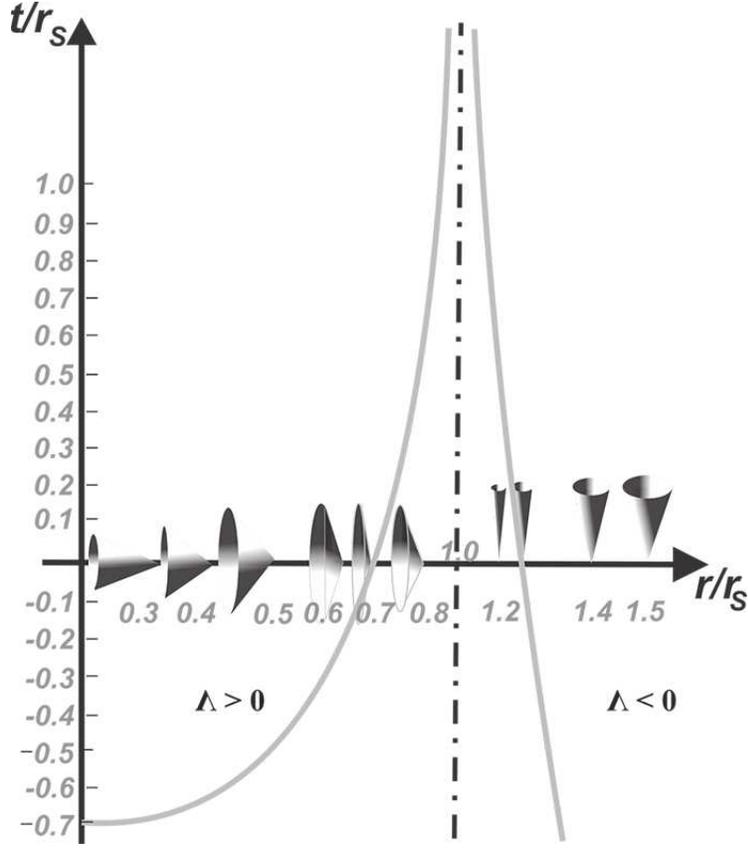}
\end{center}
\caption{This figure shows the classical (causal) structure of black hole (BH) with event horizon\cite{18}. This structure is modified with the introduction of the causal structure of SSR\cite{3}\cite{19}. We see that the speed of light changes close to the event horizon when the gravitational field is extremely high. Thus, we expect that the speed of light, the minimum speed and the cosmological constant also acquire specific values ($c'$, $V'$ and $\Lambda'$) 
during a non-classical gravitational collapse. The formation of a gravastar due to a phase transition\cite{9} that leads to the emergence of GBEC is connected to a new structure of spacetime having a positive cosmological constant (a highly repulsive core).}
\end{figure}

\subsection{Mapping of the components $rr$ and $tt$ between both metrics}

By substituting both Eqs.(40) in the metric of Eq.(39), we get GBEC-metric, as follows: 

\begin{equation}
 dS^{2}=-A\left(1-\frac{r^2}{R^{2}_{core}}\right)c^{2}dt^{2}+
 \frac{dr^{2}}{\left(1-\frac{r^2}{R^2_{core}}\right)}. 
\end{equation}

By comparing both metrics of SSR [Eq.(41)] and GBEC [Eq.(42)] with respect to their $rr$ terms
and also their $tt$ terms, we obtain two equivalences, namely: 

\begin{equation}
\frac{1}{\left(1-\frac{V'^{2}}{v^{2}}\right)}\equiv\frac{1}{\left(1-\frac{r^2}{R^2_{core}}\right)},
\end{equation}
obtained for their $rr$ terms. 

And

\begin{equation}
\frac{1}{\left(1-\frac{V'^{2}}{v^{2}}\right)}\equiv A\left(1-\frac{r^2}{R^{2}_{core}}\right), 
\end{equation}
obtained for their $tt$ terms. 

First of all, from the equivalence in Eq.(43), we get 

\begin{equation}
\frac{V'^{2}}{v^{2}}=\frac{r^2}{R_{core}^2}.
\end{equation}

By introducing Eq.(45) into Eq.(44), we finally find

\begin{equation}
 A=\frac{1}{\left(1-\frac{V'2}{v^2}\right)^2}\equiv\frac{1}{\left(1-\frac{r^{2}}{R_{core}^{2}}\right)^2}.
\end{equation}

So, now by substituting $A$ above in Eq.(42) of the GBEC-metric, we just verify that Eq.(41) for SSR-metric inside the core is indeed equivalent to Eq.(42). So, in doing this, we obtain 

\begin{equation}
dS^{2}=-\frac{c^{2}dt^{2}}{\left(1-\frac{r^2}{R^2_{core}}\right)}+\frac{dr^{2}}
{\left(1-\frac{r^2}{R^2_{core}}\right)}\equiv -\frac{c^{2}dt^{2}}{\left(1-\frac{V'^{2}}{v^{2}}\right)}+\frac{dr^{2}}{\left(1-\frac{V'^{2}}{v^{2}}\right)}. 
 \end{equation}

\begin{figure}
\begin{center}
\includegraphics[scale=0.3]{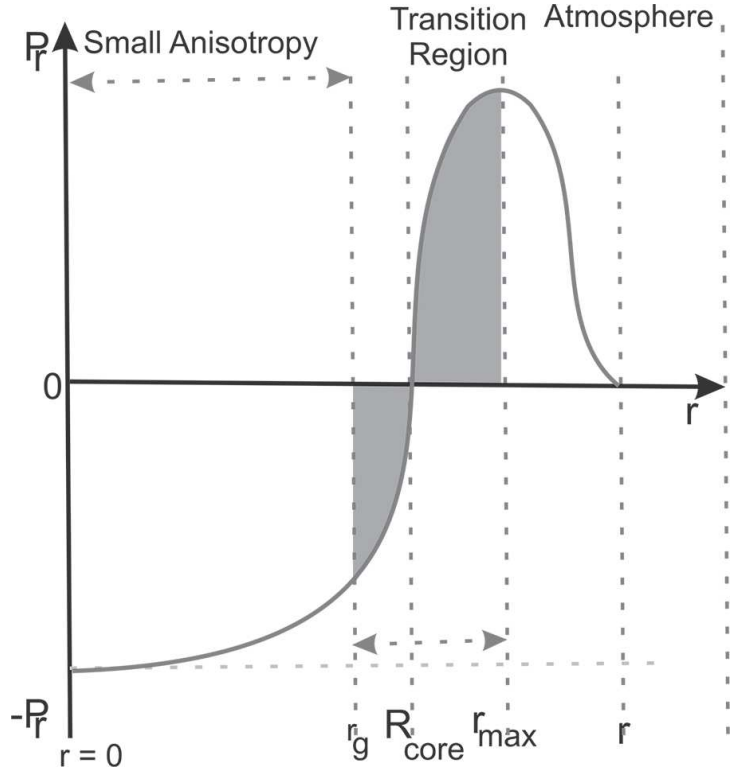}
\end{center}
\caption{Usual graph given in the literature by Cattoen\cite{9}\cite{10} for representing 
the radial pressure versus radius, showing the region inside which there is a radius of phase transition, i.e., such a radius is in somewhere in the interval $r_{g}<r<r_{max}$ to be determined. The dominant anisotropy occurs for $p(R_{core})\approx 0$\cite{9}\cite{10}, which is close to the region of baryonic matter. In this work, we show that such a transition occurs at 
a coexistence radius $R_S<R_{coexistence}<R_{core}$, where the radial pressure $p$ vanishes,
so that the singularity of event horizon is prevented. According to this figure, the pressure decreases from its maximal value at $r_{max}$ to zero at $R_{core}$ (the transition region
for Cattoen) due to the fact that gravity is reduced abruptally in this interval, where the baryonic matter is crushed into a Quark Gluon Plasma (QGP). We believe that the origin of 
such a rapid decrease of gravity comes from the vacuum anisotropy that already begins to govern the collapse for $r<r_{max}$ until reaching the phase transition (gravity/anti-gravity) with 
a null pressure at $r=R_{core}$ in the figure (actually at $r=R_{coexistence}<R_{core}$), where the anisotropy reaches its maximum value.}
\end{figure}

\section{Region of coexistence between phases}

Rewritting the Schwarzschild metric, we have 

\begin{equation}
dS^{2}=-c'^{2}dt^{2}+\frac{dr^{2}}{\left(1-\frac{2Gm}{c^{2}r}\right)}+r^{2}d\Omega^2=
-c^{2}\left(1-\frac{2Gm}{c^2 r}\right)dt^{2}+\frac{dr^{2}}{\left(1-\frac{2Gm}{c^{2}r}\right)}+
r^{2}(d\theta^2+\sin\theta^2d\phi^2),
\end{equation}
from where we obtain an effective speed of light $c'=c\sqrt{1-2Gm/c^2r}<c$. This assumption will be better justified soon when we consider the light cone in the region of phase transition or close to the event horizon in the case of classical collapse (Fig.5)\cite{18}. 

We expect that, in the region of coexistence between the two phases (gravity or baryonic phase/anti-gravity or GBEC), i.e., for a given $R_S<R_{coexistence}<R_{core}$, the light cone becomes almost closed, so that the speed of light is reduced to 
$c'=c(R_{coex.})=c\sqrt{1-2Gm/c^2R_{coex.}}$. Thus, thanks to the minimum speed, i.e., the presence of the dark cone (Fig.1)\cite{3}, we intend to show more clearly that the event horizon is almost formed at $r=R_{coex.}$, which is slightly larger than the Schwarzschild radius ($R_S=2Gm/c^2$). 

In order to obtain the coexistence radius $R_{coex.}$, we have to admit that the minimum speed increases close to the region of phase transition in such a way that $V'(>>V)$ approaches to the speed of light $c'(<<c)$ that decreases, so that the light cone becomes almost closed,
but not exactly closed at the event horizon as occurs in the classical gravitational collapse (Fig.5)\cite{18}, by preventing the singularity of event horizon. 

To know how the minimum speed increases, we have to take into account the concept of reciprocal velocity\cite{5}\cite{uncertainty}, where we have seen that the minimum speed works like a kind of ``inverse'' (reciprocal) speed ($v_{rec}$) of the speed of light, i.e., 
$v_{rec}=cV/v=v_0^2/v$\cite{5}\cite{uncertainty} with
$\Psi(v_0)=\Psi(\sqrt{cV}=1$, such that we find $v_{rec}=V$ for $v=c$. Thus, according to 
this relation for $v_{rec}$, we can get the effective minimum speed $V'$ in the region of 
phase transition, namely $V'(R_{coex.})=cV/c(R_{coex.})=cV/c
\sqrt{1-2Gm/c^2R_{coex.}}=V/\sqrt{1-2Gm/c^2R_{coex.}}$, where $V$ is the universal
minimum speed\cite{3}. 

As $V'$ approaches to $c'$ close to the region of phase transition, both speeds become equal at $R_{coex.}$, i.e., $V'(R_{coex.})=c'(R_{coex.})$ so that we write 

\begin{equation}
c'=c\sqrt{1-\frac{2Gm}{c^2R_{coex.}}}=V'=\frac{V}{\sqrt{1-\frac{2Gm}{c^2R_{coex.}}}},
\end{equation}
from where we obtain

\begin{equation}
 R_{coex.}=\frac{2Gm}{c^2(1-\xi)}=\frac{2Gm}{c^2\left(1-\frac{V}{c}\right)}, 
\end{equation}
where $\xi(=V/c)$ is the universal dimensionless constant of fine adjustment\cite{3}. And
it was already shown that there is a direct relationship between the minimum speed $V$ and the 
minimum length of quantum gravity (Planck length $L_P$), i.e.,
$V\propto L_P(=\sqrt{G\hbar/c^3})$\cite{3}. 

Eq.(50) shows the expected result by indicating that the event horizon is not formed, since we can see that the radius $R_{coex.}$ is in fact slightly larger than the Schwarzschild radius ($R_S=2Gm/c^2)$ due to the universal minimum speed $V\sim 10^{-14}$m/s that has origin in 
a quantum gravity as $V\propto L_P$. Thus a quantum gravity is responsible for preventing 
the singularity of event horizon and so it is also responsible for the existence of the 
vacuum energy/dark energy. 

We find $R_S/R_{coex.}=(1-\xi)<1$, that is to say $R_{coex.}>R_S$ by preventing the event horizon. But, if we make $\xi=0$ ($V=0$), we recover the classical case of singularity at the Schwarzschild radius (no phase transition), thus leading to the black hole (BH). 

In view of this quantum gravity effect given by the miminum speed connected to the 
cosmological constant $\Lambda$\cite{3}, we realize that the metric in Eq.(48) cannot diverge for the baryonic phase of the star, since its minimum radius (of matter) is now
$R_{coex.}>R_S$, so that the divergence of the Schwarzschild metric (Eq.(48)) is prevented. Therefore, the divergence of the metric at $R_S$ is replaced by a too high value, being 
still finite. In order to obtain such a finite result, we just substitute Eq.(50) into Eq.(48), and so we find the metric in the region of coexistence of phases for $r=R_{coex.}$, namely: 

\begin{equation}
dS^{2}_{coex.}=-\xi c^{2}dt^{2}+\frac{1}{\xi}dr^{2}+r^{2}d\Omega^{2}=-v_0^2dt^{2}+\frac{c}{V}dr^{2}+r^{2}d\Omega^{2},
\end{equation}
where $1/\xi=c/V\sim 10^{22}$ is a too large pure number, and $v_0=\sqrt{cV}$ represents a universal speed that provides the transition from gravity to anti-gravity in the cosmological scenario\cite{3}\cite{Rodrigo}. So it is interesting to note that such a speed $v_0$ also plays the role of an order parameter obtained just in the region of phase transition of a
non-classical gravitational collapse. This connection between the cosmological scenario with 
anti-gravity\cite{Rodrigo} and the phase of a repulsive core inside a gravastar by means of the same universal order parameter of transition given by $v_0$ seems to be a holographic aspect of spacetime. We will explore deeply this issue elsewhere. 

Dirac has already called attention to the importance of the well-known {\it Large Number Hypothesis} (LNH) even before the obtaining of $\xi$\cite{3}. So a given infinite greatness that appears in Physics could be removed by a more fundamental principle. In view of this, it is interesting to notice that the metric in Eq.(51) shows us that the tiny pure number $\xi$ in the denominator of the spatial term $dr$ prevents its singularity and thus also prevents an interval $dS$ of pure space-like as occurs at the event horizon of a BH (Fig.5), because the light cone does not become completely closed in the region of phase transition given by the metric in Eq.(51), since the temporal term of the metric above does not vanish
($c'(R_{coex.})=\sqrt{\xi}c=v_0$), i.e., $\xi c^{2}dt^{2}=Vcdt^{2}\neq 0$ as 
$v_0=\sqrt{cV}\neq 0$, which is exactly the order parameter of transition that indicates the begining of a new phase of anti-gravity for $r<R_{coex.}$. 

It is also interesting to note that a signal could be transmitted with speed $c'=v_0$ in the region of phase transition ($R_{coex.}$), which does not occur at the event horizon of BH, where the light cone is completely closed so that $c'=0$ (Fig.5) for $r=R_S$ (no signal).

In any way, it is important to realize that such collapsed structure (dark energy star with a thin shell of baryonic matter) can emit radiation, since the temporal term of the
metric in the phase transition region [Eq.(51)] is non zero, i.e., $-\xi c^2 dt^2$, which indicates that there is no event horizon. 

We sill realize that the Buchdahl-Bondi relation for preventing the event horizon is now better justified by the constant $\xi=V/c$, since we find

\begin{equation}
\frac{8}{9}<\frac{2Gm}{R_cc^{2}}=(1-\xi)<1, ~~ \Delta(R_c)=\frac{1}{4}\frac{(1-\xi)}{\xi}\approx\frac{1}{4\xi}>>0,
\end{equation}
where $\xi\sim 10^{-22}$\cite{3} and thus the anisotropy $\Delta=\frac{p_{t}-p_{r}}{\rho}$ is in fact so large at $R_{coex.}$, i.e., $\Delta(R_{coex.})>>0$ as already expected. We should remember that $p_t$ is the tangencial pressure and $p_r$ is the radial pressure. 

In the isotropic case, we find $p=p_{r}=p_{t}$, however the works of CFV and Debenedicts et al\cite{9}\cite{10}\cite{11}\cite{13} has demonstrated the relevance of the tangencial (transverse) pressure for a gravastar in spite of they were not able to provide a satisfactory explanation for the anisotropy $\Delta$, which is now well-understood by the new structure of GBEC with infinite vibrational modes of vacuum [Eq.(35)]. 

\section{Conclusions and Remarks}

This work establishes the connection between the spacetime with an invariant minimum speed,
i.e., the so-called Symmetrical Special Relativity (SSR) with Lorentz violation and the
Gravitational Bose Einstein Condensate (GBEC) as the central core of a star of gravitational vacuum (gravastar/dark energy star). So it was introduced a new causal structure of spacetime that reveals the existence of a vacuum inside the core with various vibrational modes, which naturally explain the well-known anisotropy that leads to the phase transition
(gravity/anti-gravity) at the coexistence radius ($R_S<R_{coex.}<R_{core}$).

The model eliminates the formation of a singularity of event horizon in an simple way
and leads to the emission of radiation by means of a phase transition between gravity and antigravity before reaching the Schwarzschild radius ($R_S<R_{coex.}$).

This fundamental mechanism for eliminating the singularity of the event horizon can open a window and new interesting perspectives on the study of preventing of physical (central) singularities of black holes replaced by black hole mimickers like the present model of dark energy star. 

The information paradox and other related issues will be also investigated within this new causal structure of spacetime with an invariant minimum speed. \\

{\noindent\bf  Acknowledgements}

The first author of this long research on SSR since 1988 dedicates his present work to the memory of Albert Einstein and Stephen Hawking who searched for understanding the true quantum black hole, whose foundations are provided by the spacetime with an invariant minimum speed as the kinematic origin of an anti-gravity given by the dark energy within a new quantum gravity scenario. In sum, SSR shows that 
classical black holes are not formed during a gravitational collapse. Thus, Einstein was right.

\end{document}